\documentclass[a4paper,11pt]{article}
\pdfoutput=1 

\usepackage{jcappub} 
\usepackage{graphicx}
\usepackage{subcaption}
\usepackage[labelformat=parens,labelsep=quad,skip=3pt]{caption}

\title{\boldmath Constraining the dark matter interpretation of the positron excess with $\gamma$-ray data}

\author[a]{Haoxiang. Zhan,\note{Corresponding author.}}

\affiliation[a]{Imperial College London,\\ Imperial College London, South Kensington Campus, London SW7 2AZ, United Kindom}

\emailAdd{haoxiang.zhan22@imperial.ac.uk}

\abstract{The particle origin of dark matter (DM) is still one of the main puzzles in modern physics. One of the most promising search strategy to detect DM at laboratories is through the indirect search of cosmic particles that are produced from DM annihilation in space. In particular, the flux of cosmic positrons has been measured with high precision by the AMS-02 experiment demonstrating that an excess above 10 GeV, with respect to the secondary production, is present. We study in this paper the possible DM origin of the positron excess finding the values of the DM mass $M$ and annihilation cross section $\langle \sigma v \rangle$ that are needed to fit high-energy positron data. In particular, we find that for DM annihilating into $b\bar{b}$ it is required to have $M=43$ TeV and $\langle \sigma v \rangle = 10^{-21}$ cm$^3$/s while for $\tau^+\tau^-$ $M=2$ TeV and $\langle \sigma v \rangle = 3\times 10^{-23}$ cm$^3$/s. If DM produce positrons, they are expected to generate gamma rays from the center of the Milky Way and around dwarf galaxy satellites of the Galaxy. We thus combine the values for the DM mass and annihilation cross section obtained with the fit to AMS-02 positron data with the upper limits derived with the non-detection of $\gamma$ rays with HESS in the direction of the Galactic center and Fermi-LAT for the combined analysis of dwarf galaxies.
The main result of the paper is that only DM annihilating into $\mu^+ \mu^-$ with a mass around 500 GeV and $\langle \sigma v \rangle = 4\times 10^{-24}$ cm$^3$/s can fit AMS-02 data and be compatible with the upper limits found with $\gamma$ rays.
As for the $\tau^+ \tau^-$ ($b\bar{b}$) channel, DM can contribute at most at a few tens $\%$ (a few \%) level.}

\begin{document}
\maketitle
\flushbottom

\section{Introduction}

Several evidences have been found about the presence of dark matter (DM) in the Universe in different astrophysical sources from clusters of galaxies up to individual galaxies \cite{pecontal2009review}.
The most precise measurement for the amount of DM in the Universe comes from the anisotropy of the Cosmic Microwave background \cite{aghanim2020planck}.
These gravitational measurements of DM can be explained in the so-called $\Lambda$CDM model which predicts that about $24\%$ of the Universe is made of DM and that the $85\%$ of the matter is made of these elusive particles. 
DM can be made of a new particle that is weakly interacting with ordinary matter, massive and non-relativistic in the current Universe, stable on a cosmological scale electrically neutral and non-baryonic. 
Weakly Interacting Massive Particles (WIMPs) were theorized a few decades ago as DM candidates with all the right properties that match the observations \cite{agrawal2015wimps}.
Different beyond the Standard Model theories of particle physics such as Supersymmetry (SUSY) and Extra dimensions have at least one particle which can be a WIMP. For example, in SUSY, the lightest neutralinos are particles stabilized by an $R$-parity symmetry and are among the most studied candidates \cite{roszkowski2018wimp,graesser2011asymmetric}. 
WIMPs are thought to be stable and weakly interacting, meaning they do not interact strongly with electromagnetic radiation, making them difficult to detect directly. 
The search for WIMPs is performed using mainly three strategies which are direct detection experiments \cite{schumann2019direct}, indirect detection experiments \cite{klasen2015indirect}, and colliders \cite{buchmueller2017search}. Despite the lack of direct evidence for WIMPs, they remain a compelling explanation for DM even if other particle candidates such as sterile neutrinos and axions are not excluded.

Indirect DM detection aims at searching for the products of DM annihilation or decay in space, such as gamma rays, neutrinos and the rarest cosmic rays (CRs) such as positrons, antiprotons and antinuclei \cite{gaskins2016review}. The rational behind indirect detection is that DM particles in space could annihilate, producing observable signals in the flux of various cosmic particles \cite{buckley2010dark,albert2020combined,kaplinghat2015galactic}. 
Typically, the cosmic particles used for indirect detection are the rarest ones such as antiprotons, gamma rays, positrons, neutrino and anti-nuclei. In fact, the more abundant CRs, such as protons and helium are produced almost entirely by supernova remnants.

Gamma rays and neutrinos are among the most promising cosmic particles for the indirect DM search because they have the advantage of traveling straight since they are not affected by magnetic fields. This permits searching for a DM signal toward the astrophysical sources with the highest predicted DM density. 
For example, the Fermi Gamma-Ray Space Telescope has searched for gamma-ray emissions from DM annihilation in the center of the Milky Way, which is the direction with the predicted highest density of DM in the Universe, and other galaxies \cite{hooper2011dark,berlin2014stringent,geringer2012dark}
Milky Way dwarf spheroidal galaxies (dSphs) are an example of an object class with a mass-to-luminosity ratio larger than 10 and thus very promising for detecting a DM signal. All the searches in Fermi-LAT data have brought no firm detection and tight constraints on the DM annihilation cross section or decay time have been published \cite{bertoni2015examining,bringmann2012fermi,albert2017searching}.
IceCube Neutrino Observatory has looked for high-energy neutrinos from DM annihilation in the Sun, from the Galactic center and with an astrophysical origin \cite{baum2017dark}.

Indirect detection has the advantage of searching DM particles in the cosmo where we know that DM exists because we observe its gravitational effects. However, this search technique is subject to various astrophysical uncertainties and background signals, making the interpretation of the results challenging (see, e.g, \cite{DiMauro:2021raz}).
In fact, cosmic particles are also produced by known astrophysical processes such as Galactic and extragalactic sources and secondary production, which is due to the collision of CRs against the atoms of the interstellar medium. Nevertheless, ongoing and future experiments continue to improve their sensitivity and provide new constraints on the nature of DM.

Recently, data of exceptional quality on the positron flux or fraction has been collected by several experiments such as PAMELA \cite{2019BRASP..83..974M} and AMS-02 \cite{aguilar2021alpha,aguilar2019towards,aguilar2024towards}. One of the most astonishing discoveries is an unexplained positron excess above $10$ GeV reported with very good precision by the AMS-02 Collaboration. The flux of positrons measured by AMS-02 is much higher than predictions for the secondary production. This result can be explained if a primary component of positrons exists. Among primary sources, self-annihilation of DM in the Galactic halo is one of the possible origins of the positron excess \cite{buckley2010dark,arguelles20211dark,stoehr2003dark}. However, possible astrophysical sources which can generate such a rise in the positron fraction can be considered: while the most widely known example includes Pulsar Wind Nebulae (PWNe) or Supernova Remnants (SNRs) \cite{reynolds2012magnetic,bucciantini2011modelling,Cholis:2018izy,xi2019gev,tang2019positron,Orusa:2021tts}.

A mechanism known as spin-down emission is thought to be responsible for PWNe emitting a significant flux of electrons and positrons. Pulsars are believed to produce intense electric fields that tear particles apart from their surfaces due to their rotating magnetic fields \cite{shen1970pulsars,zhang2001cosmic,amato2014theory}. As a result of the acceleration of these charged particles, a cascade of electromagnetic waves is induced through the emission of curvature radiation. This subsequently produces electrons and positrons. Since this emission is considered to be relatively fast and the resulting energy emission of the pulsar is negligible, a mature pulsar can be treated as a burst-like source of positrons with high energy \cite{cheng1986energetic,rees1974origin}.

The self-annihilation of DM is a hypothetical process in which DM particles collide and annihilate, producing other particles such as photons, neutrinos, and positrons \cite{cumberbatch2007local,porter2011dark}. This process occurs because the mass-energy equivalence of the DM particles is not conserved during the collision, allowing for the creation of new particles \cite{arkani2009theory}. It is believed that this process contributes to the observed CR positron excess in the Milky Way galaxy and in other galaxies \cite{crocker2010radio,bergstrom2013new}. In the paper, we investigate a possible DM origin of the positron excess by using the constraints for the propagation parameters and the DM density distribution using the method described in \cite{cirelli2011pppc}. In particular, we use the predictions for the secondary production to obtain the best-fit values of the DM mass and annihilation cross-section for different annihilation channels. Electrons and positrons produced from DM can make gamma rays through inverse Compton scattering against the low-energy photons of the interstellar radiation field. Therefore, we compare the results for the fit to the AMS-02 positron data with the constraints coming from the non-detection of gamma rays from DM coming from the FERMI-LAT and HESS observations of dSphs 
 \cite{abdallah2020search,ackermann2014dark}. A mixed channel of DM is also considered to produce a better fit.

The structure of the paper is as followed: In section \ref{model} we explain the different sources of positrons from secondary emission and DM. In section \ref{result}, the fit to AMS-02 data with DM model in different channels is presented. The fitted values for cross-section and mass of DM are compared with the data obtained from HESS project. DM fit with the mixed channel is also presented. Finally in section \ref{conclution}, we draw our conclusions.

\section{Model} \label{model}
In this section, we explain the detail of the model we are going to use for our calculations.
In particular, in Sec.~\ref{sec:prop} we report how we deal with propagation, in Sec.~\ref{sec:posdm} we write the setup for the production of $e^{\pm}$ from DM, and finally in Sec.~\ref{sec:secondary} we display how we calculate the secondary production.

\subsection{Propagation of positrons in the Galaxy}
\label{sec:prop}

The propagation of positrons in the Milky Way is commonly described by a stationary two-zone diffusion model with cylindrical boundary conditions. The number density of particles per
unit energy, $f(t, \vec{x}, E)=d N_{e^{\pm}} / d E$, satisfies the following transport equation \cite{cirelli2011pppc}:
\begin{equation}
\frac{\partial f}{\partial t}-\nabla(\mathcal{K}(E, \vec{x}) \nabla f)-\frac{\partial}{\partial E}(b(E, \vec{x}) f)=Q(E, \vec{x}) .
\label{eq1}
\end{equation}
$\mathcal{K}(E, \vec{x})$ is the diffusion coefficient function, which describes the transport through the irregularities of the random and regular magnetic fields in the Galaxy. The function $b(E, \vec{x})$ takes into account the energy losses. In particular, we include the energy losses for inverse Compton scattering and synchrotron emission. We ignore energy loss due to the scattering against ISM atoms, such as bremsstrahlung, ionization, and Coulombian interactions. In fact, these energy losses dominate only for energies below a few GeV, which is not how energy range of interest. 
In this problem, Eq.~\ref{eq1} is solved assuming a diffusive region that resembles a solid flat cylinder that fits horizontally in the Galactic plane. This region has a vertical dimension of $2L$, and a Galactocentric radius of $R = 20$ kpc. The boundary condition for $f(t, \vec{x}, E)=d N_{e^{\pm}} / d E$ is that $f$ vanishes at the edge of the cylinder. The location of the solar system is at $\vec{x} = (8.3 \, \rm{kpc},0)$.

In principle, the diffusion coefficient $\mathcal{K}$ depends on energy and position in the Galaxy. This is due to the fact that the distribution of the diffusive irregularities of the magnetic field varies throughout the Galactic halo. But taking into account the CR propagation an inhomogeneous diffusion coefficient is problematic. First, we know little about the Galactic geometry which provides information on variations of the magnetic field between 2 sizes of Galactic arms and disks. Second, it is technically difficult to compute the CR propagation if $\mathcal{K}$ varies for both energy and position.
Moreover, we are interested in electrons and positrons for energies above 10 GeV which do not travel so much in the Galaxy and thus they probably do not experience very different strengths of diffusion.
We assume that $\mathcal{K}(E, \vec{x})$ is only a function of energy, i.e.~the diffusion is homogeneous in the Galaxy ($\mathcal{K}(E, \vec{x})=\mathcal{K}(E)$).
We parameterized $\mathcal{K}(E)$ as a power law:
\begin{equation}
K(E)=\beta K_0(R / 1 \mathrm{GV})^\delta ,
\label{eq11}
\end{equation}
where $K_0$ is the normalization of the diffusion coefficient and $\delta$ is the slope. By fitting AMS-02 data for protons, helium and boron-to-carbon ratio, values of $K_0$ and $\delta$ are found to be about 0.0112 $\mathrm{kpc}^2 \mathrm{Myr}^{-1}$ and 0.70 respectively \cite{cirelli2011pppc}.  

The energy losses should be a function of energy and position in the Galaxy. However, we proceed as we did for the diffusion coefficient and assume that the function $b(E,\vec{x})$ is only a function of energy. This is a good approximation because very-high-energy $e^{\pm}$ travel only a few kpc in the Galaxy. Therefore, we take the energy losses as homogeneous in the Galaxy as done in several other papers in the past (see, e.g., \cite{Di_Mauro_2021}).
For the energy losses we use a power-law shape in energy as $b(E) = b_0 \cdot E^{\alpha}$ where the parameters are $b_0= 10^{-19}$ GeV/s and $\alpha = 1.9$. As demonstrated in Ref.~\cite{Di_Mauro_2021} the correct calculation of the energy losses taking into account inverse Compton scattering and synchrotron radiation follows approximately a power-law shape in energy with values compatible with the one we use.

\subsection{Positrons from dark matter}
\label{sec:posdm}

Positrons and electrons can also be produced as a result of the pair annihilation or decay of DM particles. The source term $\mathcal{Q}$ is defined as \cite{cirelli2011pppc}:
\begin{equation}
\begin{aligned}
\mathcal{Q}_{\mathrm{ann}}(\vec{x}, E) & =\epsilon\left(\frac{\rho(\vec{x})}{m_{D M}}\right)^2 \sum_f\langle\sigma v\rangle_f \frac{d N_{e^{\pm}}^f}{d E} \\
\mathcal{Q}_{\mathrm{dec}}(\vec{x}, E) & =\left(\frac{\rho(\vec{x})}{m_{D M}}\right) \sum_f \Gamma_f \frac{d N_{e^{\pm}}^f}{d E} ,
\end{aligned}
\label{123}
\end{equation}
where $\epsilon$ can either take the value $1/2$ or $1/4$ depending if the DM is self-conjugate. $f$ is the index that spans the possible SM particles which are products of DM decay or annihilation. $d N_{e \pm}^f / d E$ denotes the energy spectrum of $e^{\pm}$ produced by a single DM reaction process, which is usually calculated using the Pythia Monte Carlo code \cite{cirelli2011pppc}. $\rho(\vec{x})$ denotes the density of DM particles in the Milky Way halo, which we
assume to be spherically symmetric and with a radial
distribution described by the NFW profile:
\begin{equation}
\rho_{\mathrm{NFW}}(r)=\rho_s \frac{r_s}{r}\left(1+\frac{r}{r_s}\right)^{-2}
\label{2.4}
\end{equation}
DM annihilation/decay is usually considered to occur through a single channel following a model-independent analysis. Our focus is this paper is on 3 distinct channels: $\mu^{+} \mu^{-}, \tau^{+} \tau^{-}, b \bar{b}$, which take into account leptonic and hadronic channels.
We take the source spectrum using the tables released in \cite{cirelli2011pppc}. These have been produced using the Pythia Monte Carlo generator that simulates events produced from a resonance of energy equal to twice the DM mass, for annihilating DM, decaying into a couple of SM particles.
These spectra take into account electroweak correction which is relevant when the DM mass is greater than the electroweak scale.

The differential flux of $e^{\pm}$ produced from DM annihilation can be calculated as $d \Phi_{e^{\pm}} / d E=v_{e^{\pm}} f / 4 \pi$  where $f$ represents the solution of the propagation equation evaluated at the position $\vec{x}_{\odot}$ of the Solar System. In particular, the flux is written, for annihilating and decaying DM, as:
\begin{equation}
\frac{d \Phi_{e^{\pm}}}{d E}(E, \vec{x})=\frac{v_{e^{\pm}}}{4 \pi b(E, \vec{x})}\left\{\begin{array}{c}
\frac{1}{2}\left(\frac{\rho(\vec{x})}{M_{\mathrm{DM}}}\right)^2 \sum_f\langle\sigma v\rangle_f \int_E^{M_{\mathrm{DM}}} d E_{\mathrm{s}} \frac{d N_{e^{\pm}}^f}{d E}\left(E_{\mathrm{s}}\right) I\left(E, E_{\mathrm{s}}, \vec{x}\right)  \\
\quad\left(\frac{\rho(\vec{x})}{M_{\mathrm{DM}}}\right) \sum_f \Gamma_f \int_E^{M_{\mathrm{DM}} / 2} d E_{\mathrm{s}} \frac{d N_{e^{\pm}}^f}{d E}\left(E_{\mathrm{s}}\right) \mathcal{I}\left(E, E_{\mathrm{s}}, \vec{x}\right)  ,\quad 
\end{array}\right.
\end{equation}
where $E_{\mathrm{s}}$ is the energy of the positron at the location of the production (the DM halo position), $v_e$ is the $e^{\pm}$ velocity, which is approximatively equal to the velocity of light. 
$\langle\sigma v\rangle$ is the annihilation cross section averaged for the DM velocity in the Galactic halo.
$\mathcal{I}\left(E, E_{\mathrm{s}}, \vec{x}\right)$ is the generalized halo function, which encapsulates all the astrophysics and is independent of the particle physics model. For different models of DM distribution and positron propagation function, the halo function is different. halo functions provide information about the final spectrum we are interested in. $\mathcal{I}$ obeys the boundary condition that $\mathcal{I}(E, \vec{x})=1$ and $\mathcal{I}(E_s, \vec{x})=0$ on the boundary of the diffusion cylinder. 

\begin{figure}
    \centering
    \includegraphics[width=10cm]{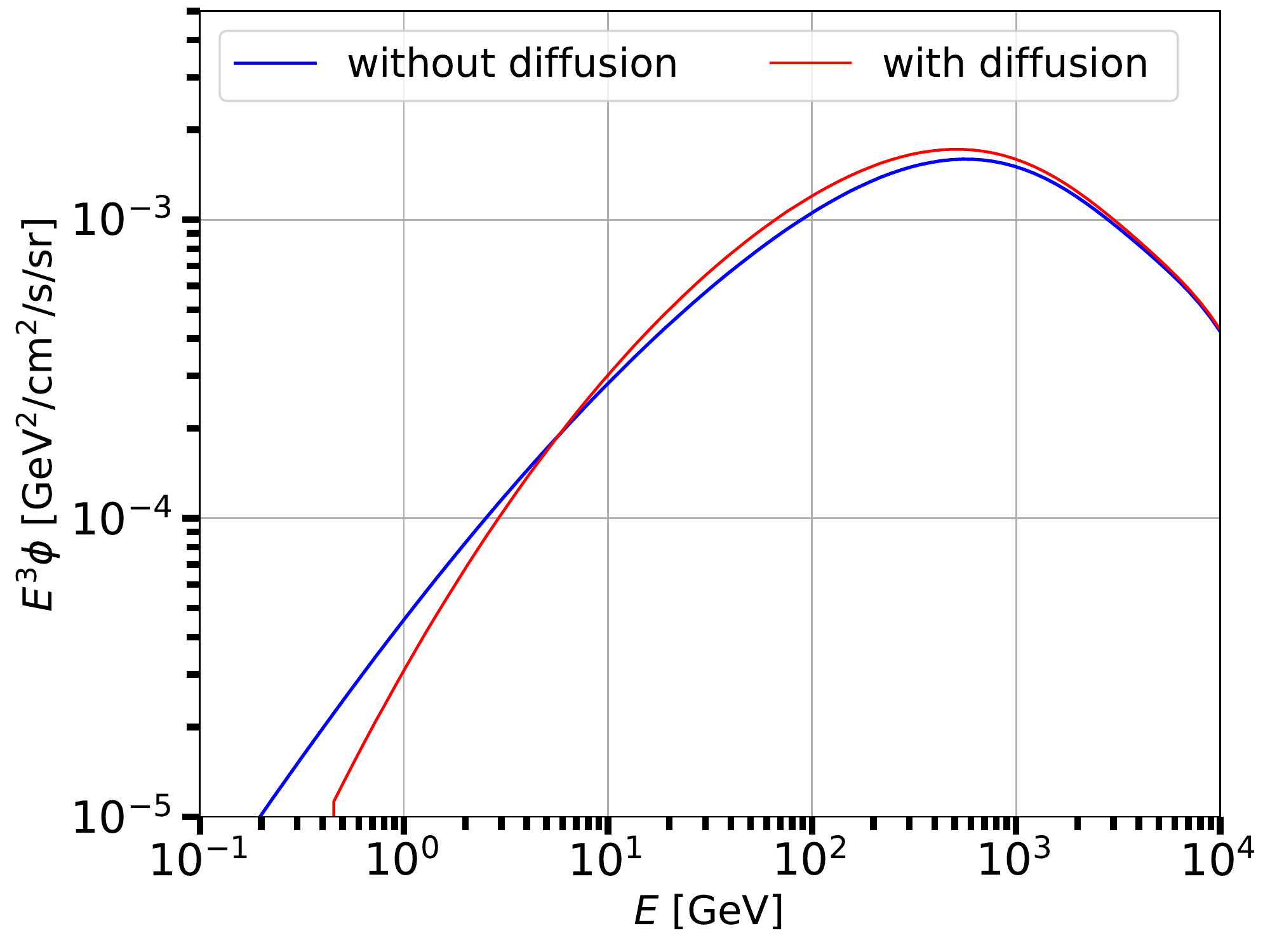}
    \caption{The positron fluxes of dark matter with and without diffusion. If diffusion is neglected, the halo function $I\left(E, E_{\mathrm{s}}, \vec{x}\right)=1$. The reduced halo function which is in the form \ref{eq22} is used in the calculation of diffusion is considered. Both lines own similar trend, the main difference is at lower energy regions between 0.2 and 5 GeV.}
    \label{fig1}
\end{figure}

A reduced halo function $\mathcal{I}\left(\lambda_D, \vec{x}\right)$ can be derived with the  assumption that both diffusion function and energy-lost term are space-independent. $K$ is in the form of  \ref{eq11} and $b(E) = b_0 \epsilon^\alpha$. The propagation scale $\lambda_D$ characterizes the CR electron horizon and depends on the energy in terms of the ratio of the diffusion coefficient to the 
energy-loss rate. $\lambda_D$ is defined as:
\begin{equation}
\lambda^2 \equiv 4 \int_E^{E_s} d E^{\prime} \frac{K\left(E^{\prime}\right)}{b\left(E^{\prime}\right)}
\end{equation}
The halo function at the location of the earth is computed numerically:
\begin{equation}
\mathcal{I}\left(\lambda_D\right)=a_0+a_1 \tanh \left(\frac{b_1-\ell}{c_1}\right)\left[a_2 \exp \left(-\frac{\left(\ell-b_2\right)^2}{c_2}\right)+a_3\right]
\label{eq22}
\end{equation}
Where $\ell=\log _{10}\left(\lambda_D / \mathrm{kpc}\right)$. Parameters in \ref{eq22} are fit parameters which can be found in Figure 7 of Ref.~\cite{cirelli2011pppc}. Parameter values vary with different DM halo functions and 3 different models "MIN", "MED" and "MAX". 
We show in Figure \ref{fig1} the positron fluxes of DM with or without considering diffusion. Both curves are similar above 10 GeV while at lower energies, between 0.2 and 5 GeV, there could be a difference that reaches even a factor of two. This plot demonstrates the importance of adding diffusion, in addition of energy losses, in the calculation of the flux of $e^{\pm}$ from DM annihilation. 
Other effects, such as reacceleration or convection could be important and are taken into account in the propagation of $e^{\pm}$. However, their effect is minimal for energies above 10 GeV.

\subsection{Secondary positrons}
\label{sec:secondary}

At low energies, most of the positrons detected by AMS-02 are probably due to the collision of CRs with the atoms of the interstellar medium (see, e.g., \cite{Delahaye:2008ua,DiMauro:2023oqx}). This process, which is called secondary production, is mainly due to the collision of CR protons and helium nuclei with the atoms of the interstellar medium which are mainly made of hydrogen and helium atoms. In proton-proton collisions (or nuclei collisions), secondary positrons and electrons are not directly produced but are rather produced by the decay of intermediate mesons and hadrons.

The source term for secondary $e^{\pm}$ is defined as the convolution between the primary CR flux $(\phi)$, the interstellar gas density $\left(n_{\mathrm{ISM}}\right)$ and the energy-differential cross section for electrons and positrons production $\left(d \sigma / d T_{e^{\pm}}\right)$. To be specific, the overall source term is computed as the sum of all the possible combinations of the $i$-th CR species with the $j$-th ISM components as:
\begin{equation}
q\left(T_{e^{\pm}}\right)=\sum_{i, j} 4 \pi n_{\mathrm{ISM}, j} \int d T_i \phi_i\left(T_i\right) \frac{d \sigma_{i j}}{d T_{e^{\pm}}}\left(T_i, T_{e^{\pm}}\right)
\end{equation}
where $T_{e^{\pm}}$ is the kinetic energy for electrons and positrons, $\phi_i$ is the incoming CR flux with the kinetic energy $T_i$ and $d \sigma_{i j} / d T_{e^{\pm}}$ is the energy-differential production cross section for the reaction $i+j \rightarrow e^{\pm}+X$.

To obtain the source term  from the production cross sections of pions
and kaons, we start by considering the channels which produce intermediate $\pi^{+}$. These channels contribute to $80-90 \%$ of secondary positron production \cite{orusa2022new}. Upon production, the majority of pions decay first into muons, followed by muons decaying into positrons. 
We calculate the secondary prediction using the production cross section as in Ref.~\cite{orusa2022new} and using the MED propagation parameters \cite{Genolini:2021doh}.

\section{Results} 
\label{result}
In this section, we report the results for the best fit of the DM parameters obtained through a fit to the positron AMS-02 data. Our model is based on the secondary production plus a primary component which is due to DM particle annihilations.
Then we will compare the best-fit values for $\langle \sigma v\rangle$ and DM mass with the constraints that come from the non-detection of $\gamma$ rays from dSphs by Fermi-LAT and from the Galactic center by HESS.

\subsection{Pure annihilation channels}
We first make the simplest assumption that DM annihilates into a single channel.
This is the most common approach done in Astroparticle Physics (see, e.g., \cite{ackermannn2014dark,Genolini:2021doh}).
In this case, DM annihilates with a branching ratio equal to one into a couple of leptons, quarks, or gauge bosons of the Standard Model. In particular, we consider the following three channels: $\mu^{+} \mu^{-}, \tau^{+} \tau^{-}, b \bar{b}$. We choose these channels as representative of leptonic and hadronic channels. In fact, the hadronic channels have all very similar source spectra and so they would all give comparable results (see, e.g. \cite{cirelli2011pppc}).

The first step is to use data on the positron flux from the AMS-02 Collaboration \cite{aguilar2024towards} to estimate the values of the DM mass and annihilation cross section. We take into account the secondary positrons originated in CR spallations since this is the most important contribution below a few tens of GeV (see, e.g., \cite{Delahaye:2008ua}). Primary positrons from sources like PWNe and SNRs are ignored in our analysis since we want to find conservative limits for DM. In fact, assuming that there is also a flux of $e^+$ from PWNe, for example, the space left for a DM contribution will reduce and thus the annihilation cross section needed to fit the AMS-02 data will be smaller with respect to what presented here. Therefore, in our paper DM would be the main contributor to the observed cosmic positrons above a few tens of GeV.

By using the flux of positrons from secondary production and DM, as explained in Sec.~\ref{sec:posdm} and \ref{sec:secondary}, we fit the AMS-02 flux data, as given by the sum of the flux from DM and secondary positrons, to the data. For the total positron flux model, 3 are the parameters that we fit: the annihilation cross-section $\langle \sigma v \rangle$, DM mass $M$, and a re-normalization factor of secondary positrons that we call $q$. The latter parameter multiplies the reference secondary flux calculated using the $e^{\pm}$ production cross sections as in \cite{orusa2022new} and the MED propagation model \cite{Genolini:2021doh}. 
This renormalization factor is justified by the uncertainties present in the production cross sections of positrons and electrons for CR interacting with ISM atoms. In Ref.~\cite{Orusa:2022pvp} these uncertainties have been estiamted to be at the level of at least 10$\%$. 
We decide to fit only positron data above 10 GeV to minimize the effect of the solar modulation, which we take into account with the force field approximation and a Fisk potential. 
Fitting of positron flux from DM self-annihilation is performed based on the Minuit package \cite{James:1975dr}. 
Electroweak corrections are included in the DM source spectra as the effect is non-negligible for DM masses above 1 TeV \cite{cirelli2011pppc}.
The fits are performed using the {\tt Minuit} minimization package \cite{James:1975dr}.
The number of data points is 47 with 3 fit parameters. 

\begin{figure}
\includegraphics[width=0.51\textwidth]{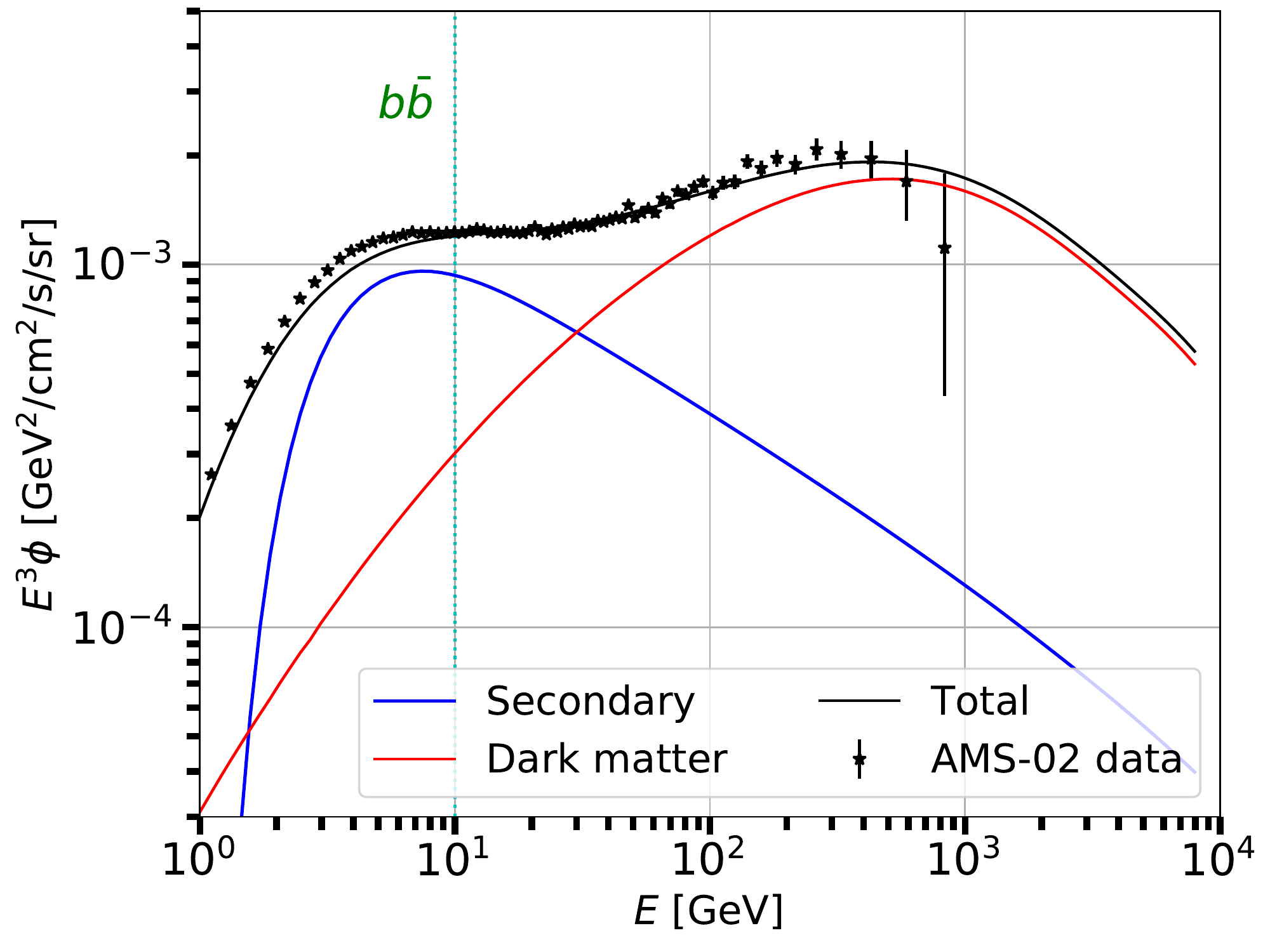}
\includegraphics[width=0.51\textwidth]{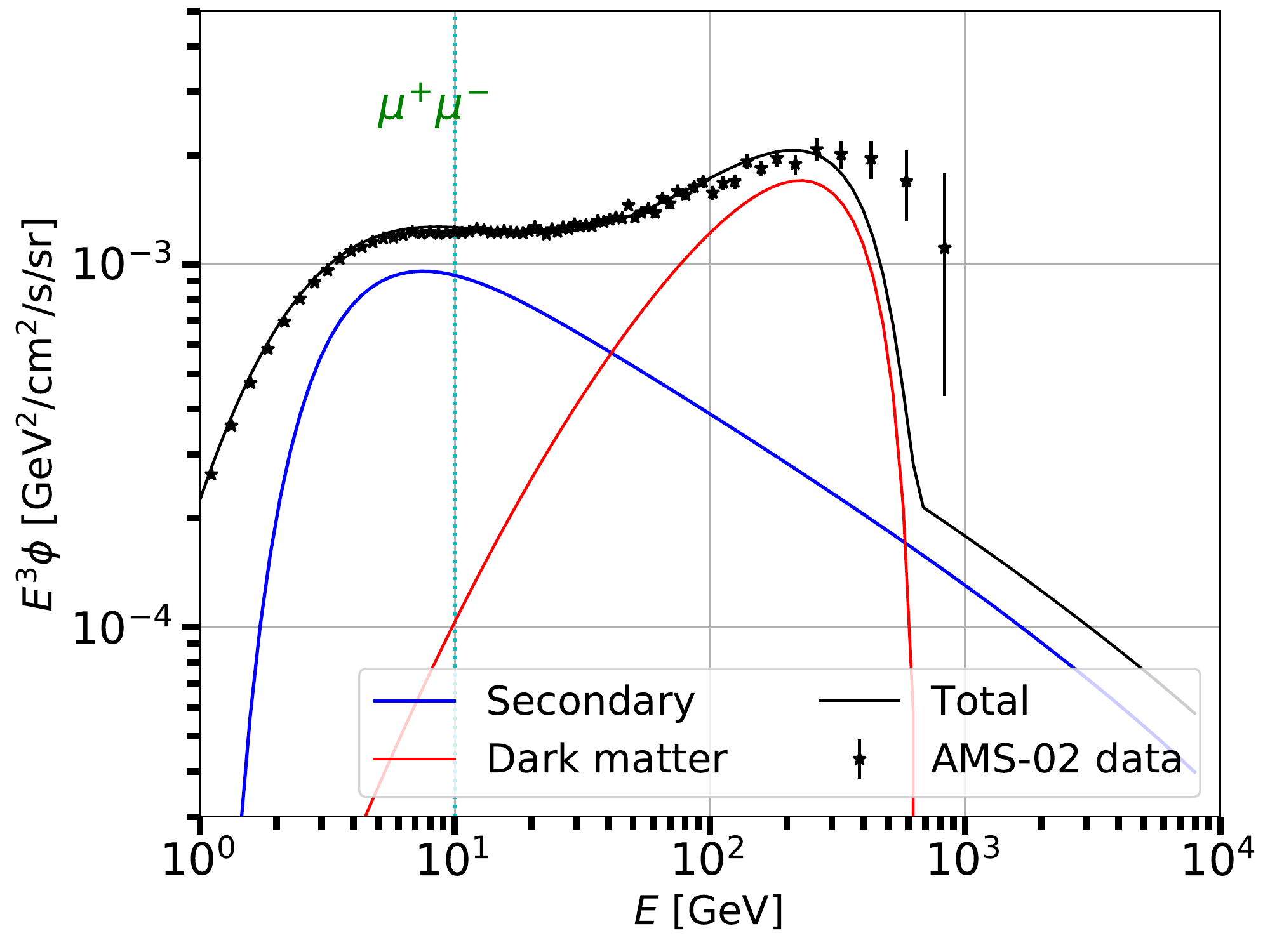}
\caption{Results of our fit to the AMS-02 positron flux data. We consider the $b \bar{b}$ (left panel) and the $\mu^{+} \mu^{-}$ (right panel) DM annihilation channels. We also show a vertical curve at 10 GeV to stress that we only fit AMS-02 data above this energy. }
\label{1234}
\end{figure}

\begin{figure}
\centering\includegraphics[width=0.54\textwidth]{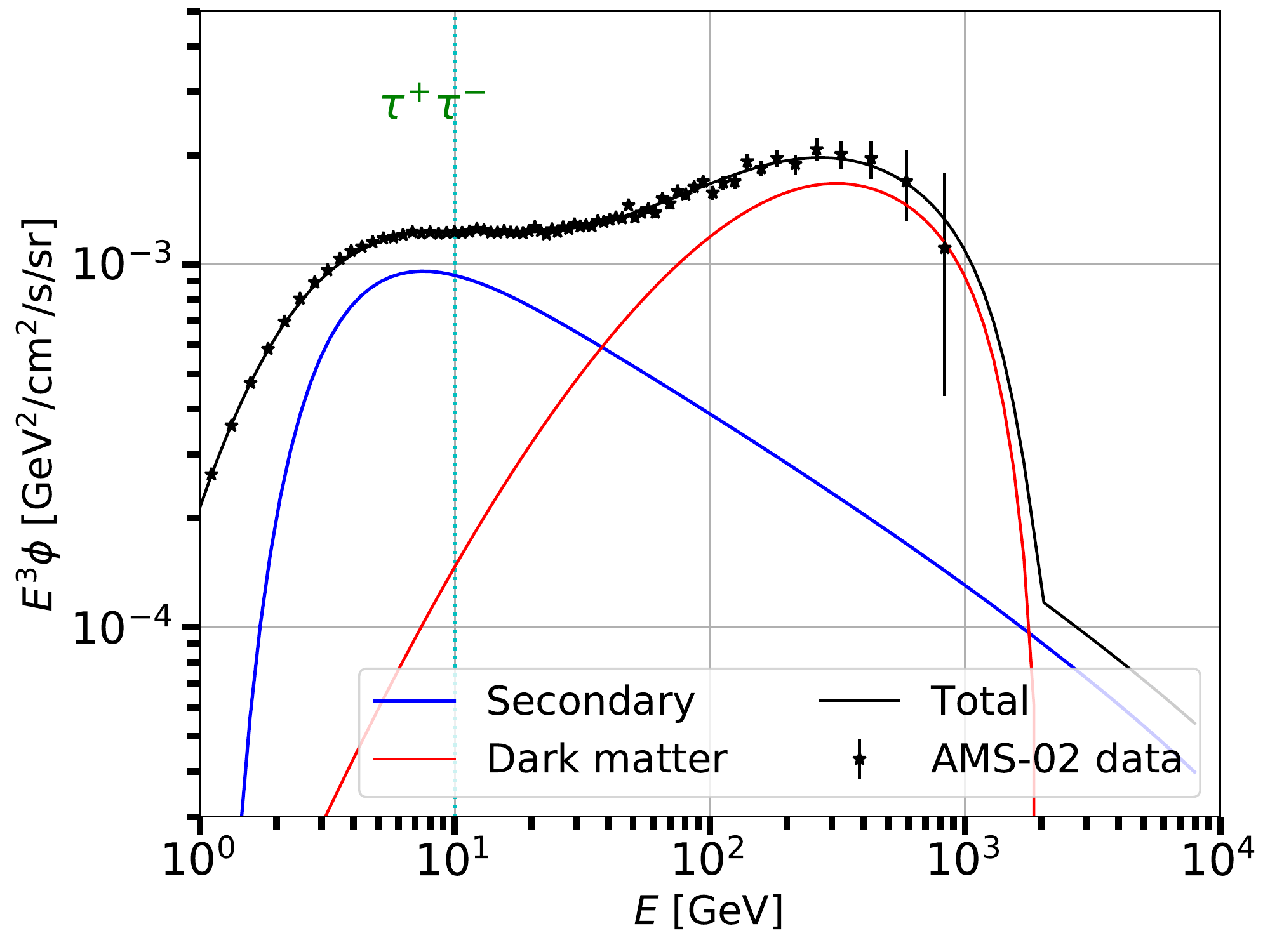}
\caption{Same as Fig.~\ref{1234} but for the $\tau^+\tau^-$ annihilation channel.}
\label{1234bis}
\end{figure}

Fig.~\ref{1234} and \ref{1234bis} show the best fit to the AMS-02 positron flux data for the three different annihilation channels considered in the paper. 
In Table.~\ref{table1} we report a summary of the results. 
The best-fit DM mass and cross-section for $b \bar{b}$ is 42.7 TeV and $9.8 \times 10^{-22}$ cm$^3$ /s with a $\chi^2$ value of 60. The best fit for $\mu^{+} \mu^{-}$ is $M = 667$ GeV and $3.98 \times 10^{-24}$ cm$^3$ /s with a $\chi^2$ value of 83 while for $\tau^{+} \tau^{-}$ is 1980 GeV and $3.16 \times 10^{-23}$ cm$^3$ /s with a $\chi^2$ value of 28.09. 
The DM mass varies enormously ranging from a few hundred GeV to a few tens of TeV. This is due to the different source spectra for the production of positrons that the different channels have. In fact, the $\mu^+ \mu^-$ annihilation channel has a much narrower spectrum with respect to the other two. This implies that the best fit mass results to be smaller. Another consequence is that the goodness of fit for the $\mu^+ \mu^-$ annihilation channel is worse than the other two because the data show a broad bump shape which cannot be reproduced by DM annihilating into $\mu^+ \mu^-$. Therefore, it fits well with the lower-energy data while failing to be compatible with the highest-energy ones. In fact, the $\tau^+ \tau^-$ channel provides a very good fit with a reduced $\chi^2$ much smaller than 1. Instead, the $\mu^{+} \mu^{-}$ and $b \bar{b}$ channels provide a sufficiently decent fit with a reduced $\chi^{2}$ between 1 and 2.
In particular, for energies between $10^2 -10^3$ GeV the case with $\mu^{+} \mu^{-}$ fails to fit well the highest AMS-02 data regardless of the large data uncertainties. 

\begin{table}[h!]
\centering
\begin{tabular}{ |p{3cm}||p{3cm}|p{3cm}|p{3cm}|  }
 \hline
 \multicolumn{4}{|c|}{Fit parameters for different channels} \\
 \hline
 Parameters/ Channels &$\tau^{+} \tau^{-}$& $b \bar{b}$ &$\mu^{+} \mu^{-}$\\
 \hline
 $M$ [GeV] & 1980 &42660& 666.8\\
 $\langle \sigma v \rangle$ [cm$^3$/s] & $3.16 \times 10^{-23}$ & $9.8 \times 10^{-22}$ & $3.98 \times 10^{-24}$\\
 $q$ &1.177&0.966& 1.253\\
 $\tilde{\chi}^2$ & 0.64&1.36&1.89\\
 
 \hline
 
\end{tabular}
\caption{Fit parameters and reduced $\chi^2$ for 3 different channels.$b \bar{b}$ and $\mu^{+} \mu^{-}$ have reasonable value for $\chi^2$ which is close to 1. Fitted DM mass varies enormously ranging from a few hundred GeV to a few ten thousand GeV. $\tau^{+} \tau^{-}$ is the least best fit channel with a reduced $\chi^2$ of 0.396.}
\label{table1}
\end{table}

\begin{figure}
\includegraphics[width=0.51\textwidth]{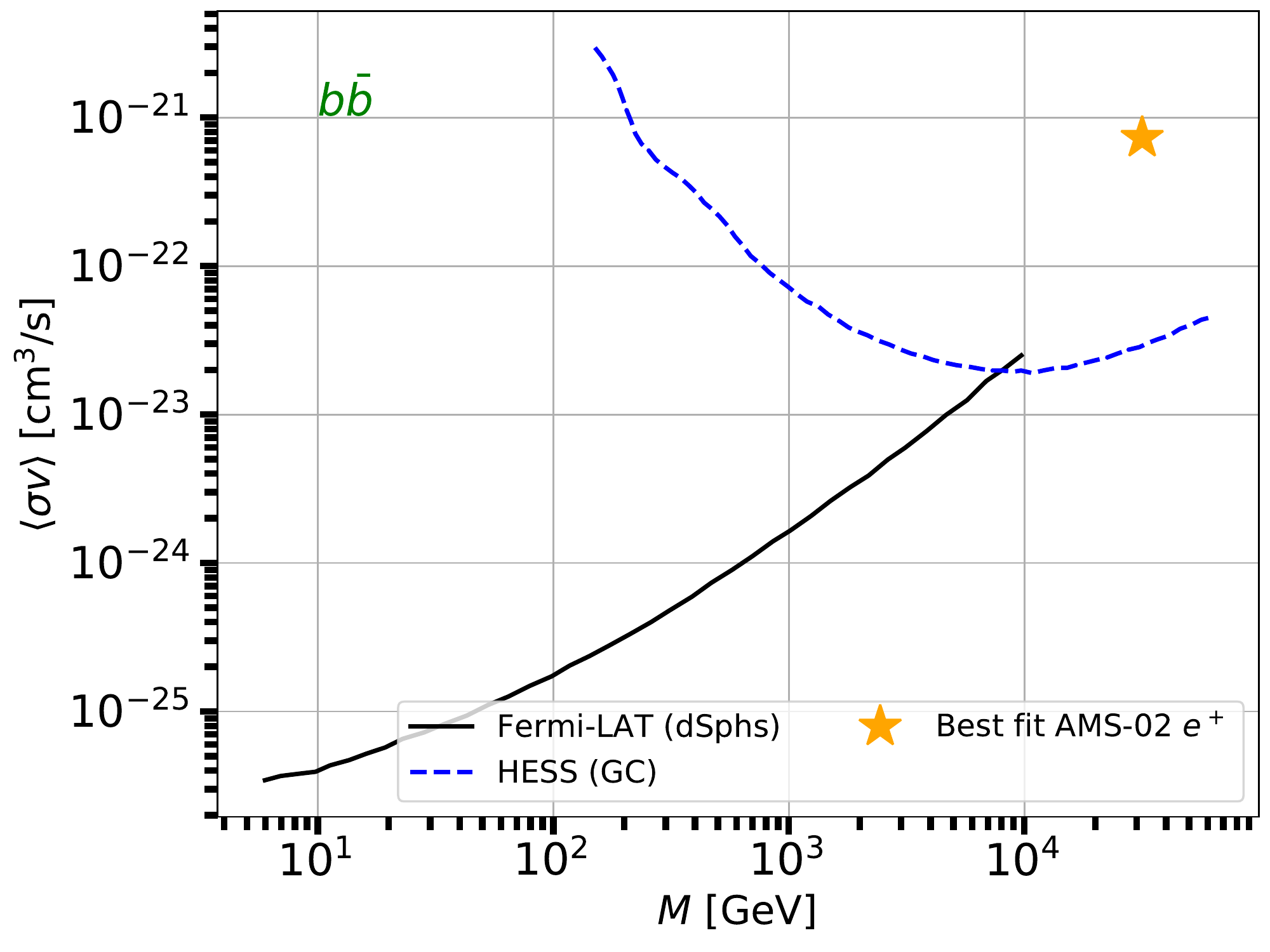}
\includegraphics[width=0.51\textwidth]{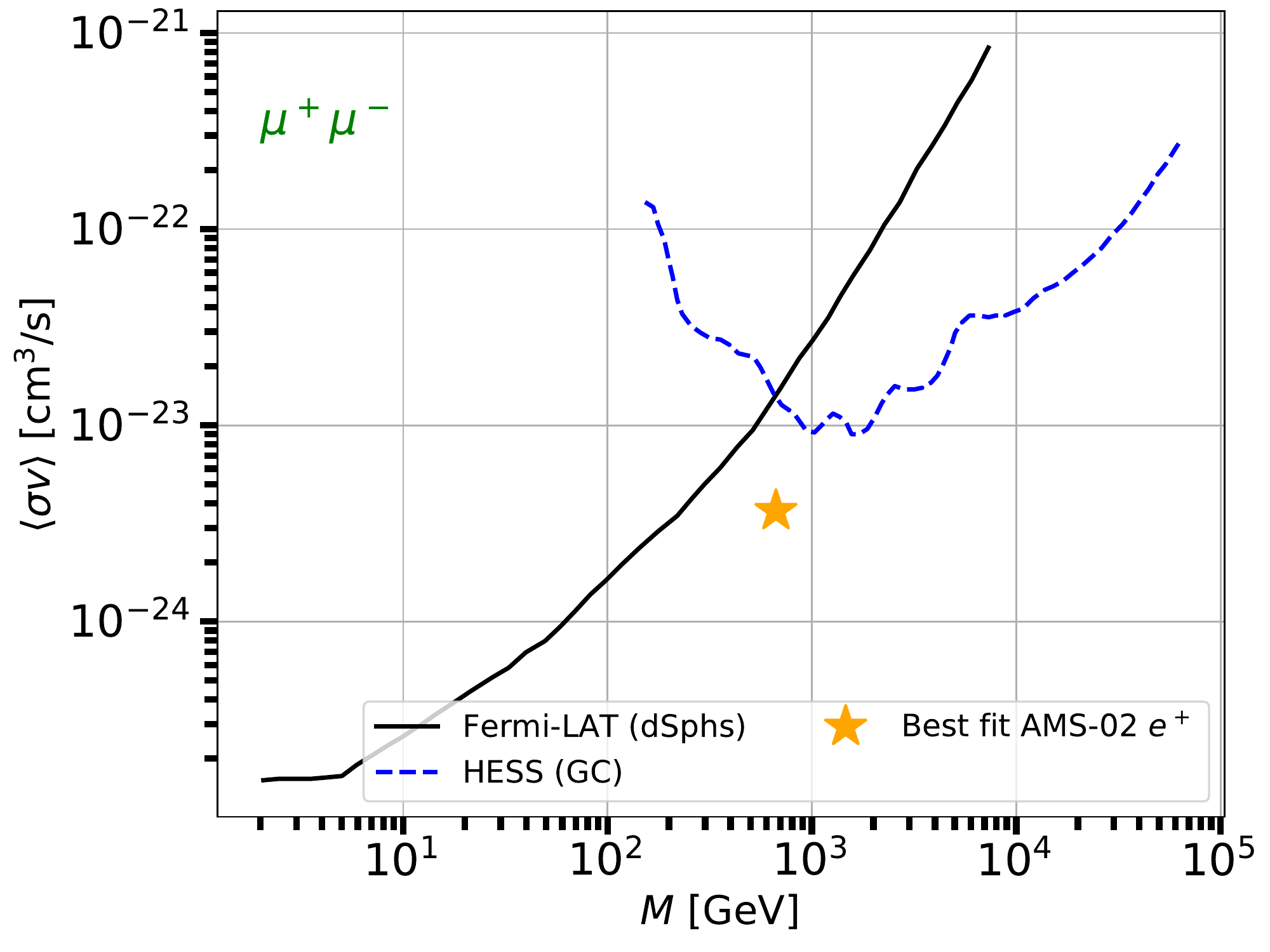}
\caption{Comparison between the DM mass and cross-section fitted to AMS-02 $e^+$ data with the upper limits found by the HESS and Fermi Collaborations \cite{abdallah2020search,ackermann2014dark}. The interpretation of the channel $b \bar{b}$ has been excluded by both HESS and Fermi derived limits on the DM cross-section. $\mu^{+} \mu^{-}$ channel is in good agreement with HESS and Fermi derived limit despite the fitted values agreeing the least with AMS-02 data. As for $\tau^{+} \tau^{-}$ channel, the HESS limit excluded our fitted value while the one for Fermi does not.  }
\label{12345}
\end{figure}

\begin{figure}
\centering\includegraphics[width=0.54\textwidth]{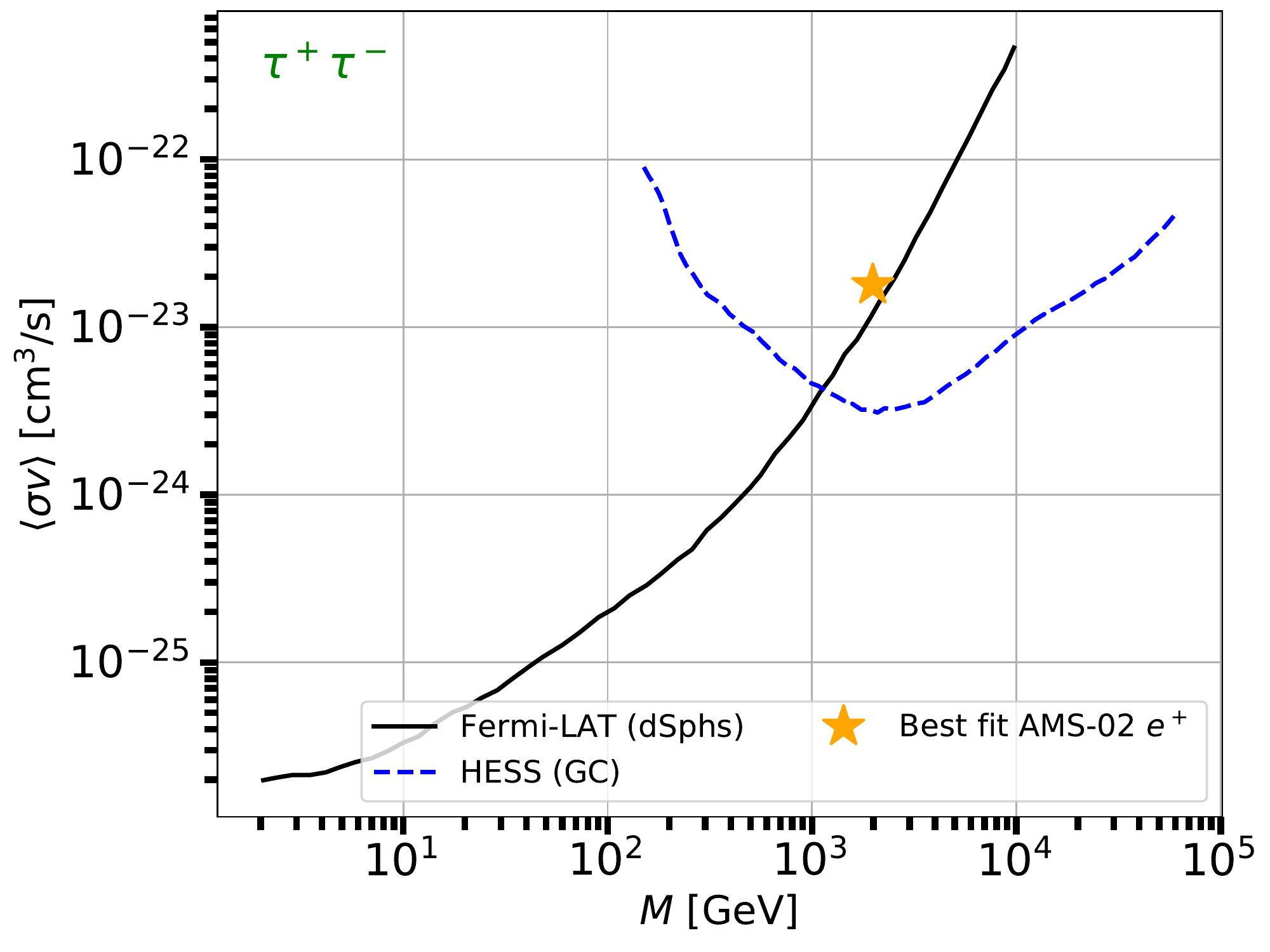}
\caption{Same as Fig.~\ref{12345} but for the $\tau^+\tau^-$ channel.}
\label{12345bis}
\end{figure}

It is thus possible that DM explains the positron excess measured by AMS-02 above 10 GeV. 
Our best case scenario is DM particles annihilating into $\tau^+\tau^-$.
However, if DM produces positrons, a flux of $\gamma$ rays is expected to be generated from different astrophysical sources.
The ones that are the most interesting to look at are the Galactic center of the Milky Way and dSphs which are very dense DM. In particular, dSphs have a mass-to-luminosity ratio of the order of 10-1000 and they have a predicted very low astrophysical production of $\gamma$ rays.
Different groups have searched for a $\gamma$-ray signal in dSphs using Fermi-LAT data (see, e.g., \cite{ackermannn2014dark}) without finding any significant detection.
Instead, Cherenkov telescope experiments have searched for a signal in the Galactic center at energies above 100 GeV, where the foreground that comes from the Galactic interstellar emission is lower (see, e.g., \cite{abdallah2020search}).
Also in this case there is no detection and thus upper limits for the annihilation cross-sections have been found.
Figs.~\ref{12345} and \ref{12345bis} outline the comparison of best-fit values for the DM mass and cross-section found with AMS-02 $e^+$ data with the upper limits derived by HESS (High Energy Stereoscopic System, an array of imaging atmospheric Cherenkov telescopes) \cite{abdallah2020search} for the Galactic center and by Fermi-LAT \cite{ackermannn2014dark} from dSphs. For the analysis of dSphs the Fermi-LAT Collaboration used several years of data and they modeled the expected gamma-ray emission from DM annihilation, estimated background emissions from various sources, and employed a likelihood analysis method to search for excess $\gamma$-ray emissions that could be attributed to DM annihilation. A joint likelihood analysis was performed to combine the results from all 25 satellite galaxies, increasing the sensitivity of the analysis. 

We can see from Figs.~\ref{12345} and \ref{12345bis} that the best-fit annihilation cross section and mass obtained with the $\mu^{+} \mu^{-}$ channel agree with the HESS and Fermi-LAT upper limits. In particular, the best-fit is a factor of about two time smaller.
Instead, the best-fit for the $\tau^{+} \tau^{-}$ and $b\bar{b}$ channels is above the $\gamma$-ray upper limits. In particular, in the case of the $b\bar{b}$ channel DM can contribute to the positron excess at most at the level of a few $\%$ while for the $\tau^{+} \tau^{-}$ at the $20\%$ level.
Therefore, the only channel that survives to the existing $\gamma$-ray upper limits, is the $\mu^{+} \mu^{-}$ one. However, this DM candidate does not provide a good fit at all AMS-02 positron data.
For the other channels this paper provides very tight constraints on a possible DM contribtion to cosmic positrons.
We stress that probably DM is not the only contributor to the very-high-enery part of the AMS-02 positron data. Therefore, the fact that the $\mu^{+} \mu^{-}$ annihilation channel does not provide a good fit does not imply that this model is excluded.

\subsection{Mixed channels}

We also test a more complicated model where we assume that DM annihilates into two channels. This implies that there is a branching ratio ($Br$) for the annihilation into a couple of leptons or quarks and $1-Br$ for a second annihilation channel.
We apply thus the same analysis procedure as before but this time there is one additional fit parameter which is the branching ratio. For example for the case where DM annihilates in $\mu^+\mu^-$ and $\tau^+\tau^-$ we use a $Br$ for DM annihilating into the former and $1-Br$ for the latter.

\begin{figure}
  \begin{subfigure}{8cm}
    \centering\includegraphics[width=8cm]{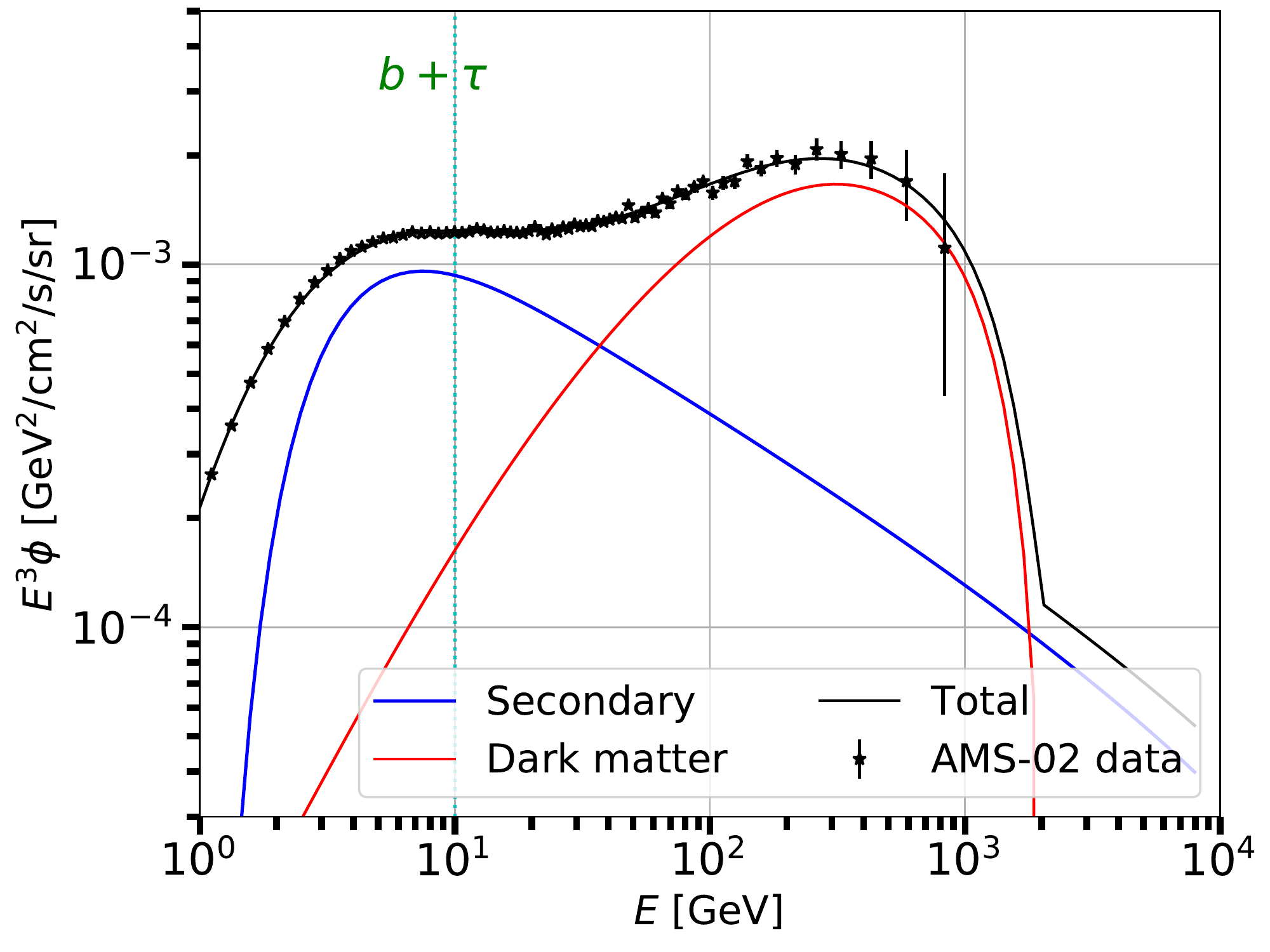}
    
  \end{subfigure}
  \begin{subfigure}{5cm}
    \centering\includegraphics[width=8cm]{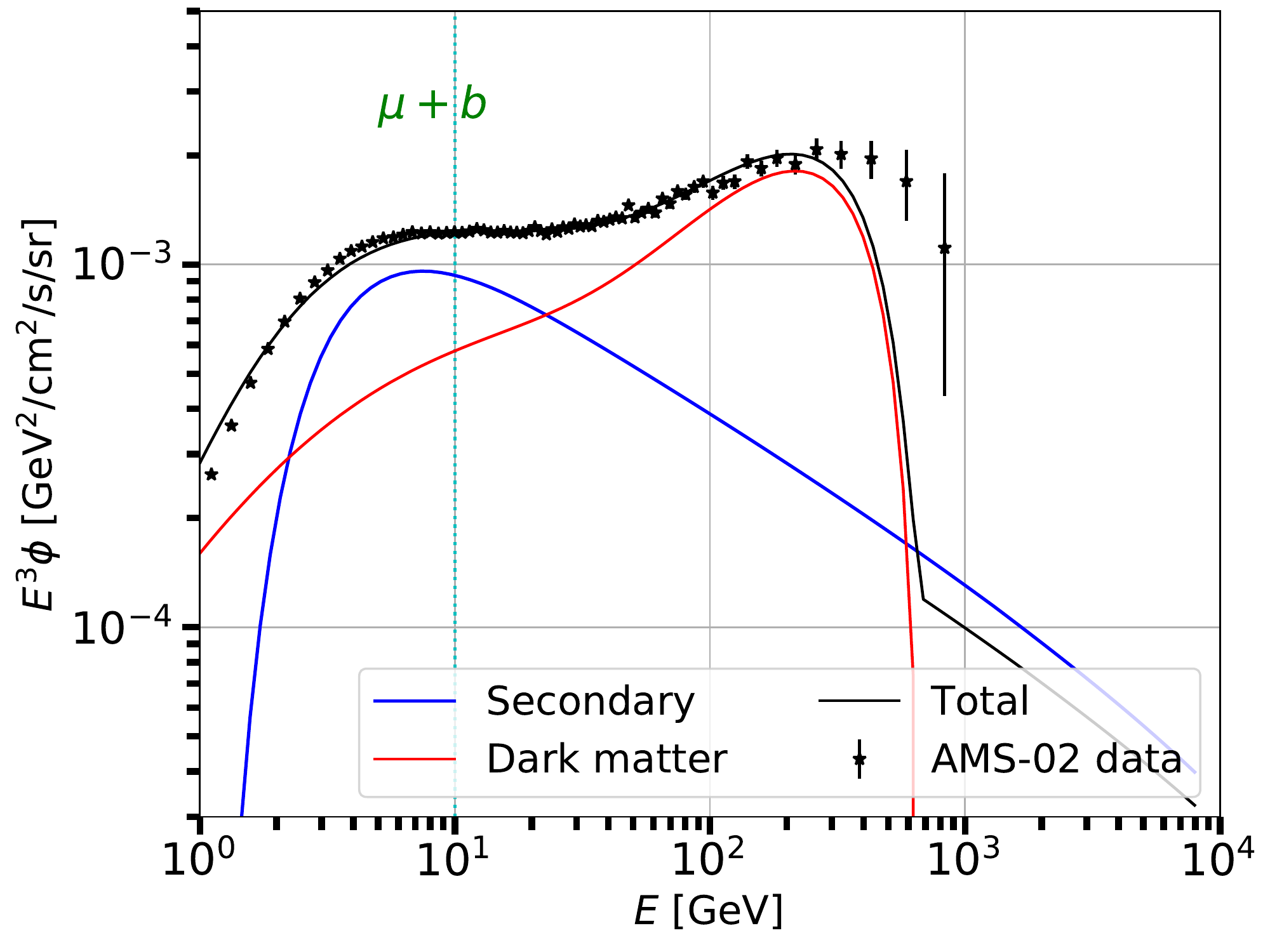}
    
  \end{subfigure}
 
  \begin{subfigure}{5cm}
    \centering\includegraphics[width=8cm]{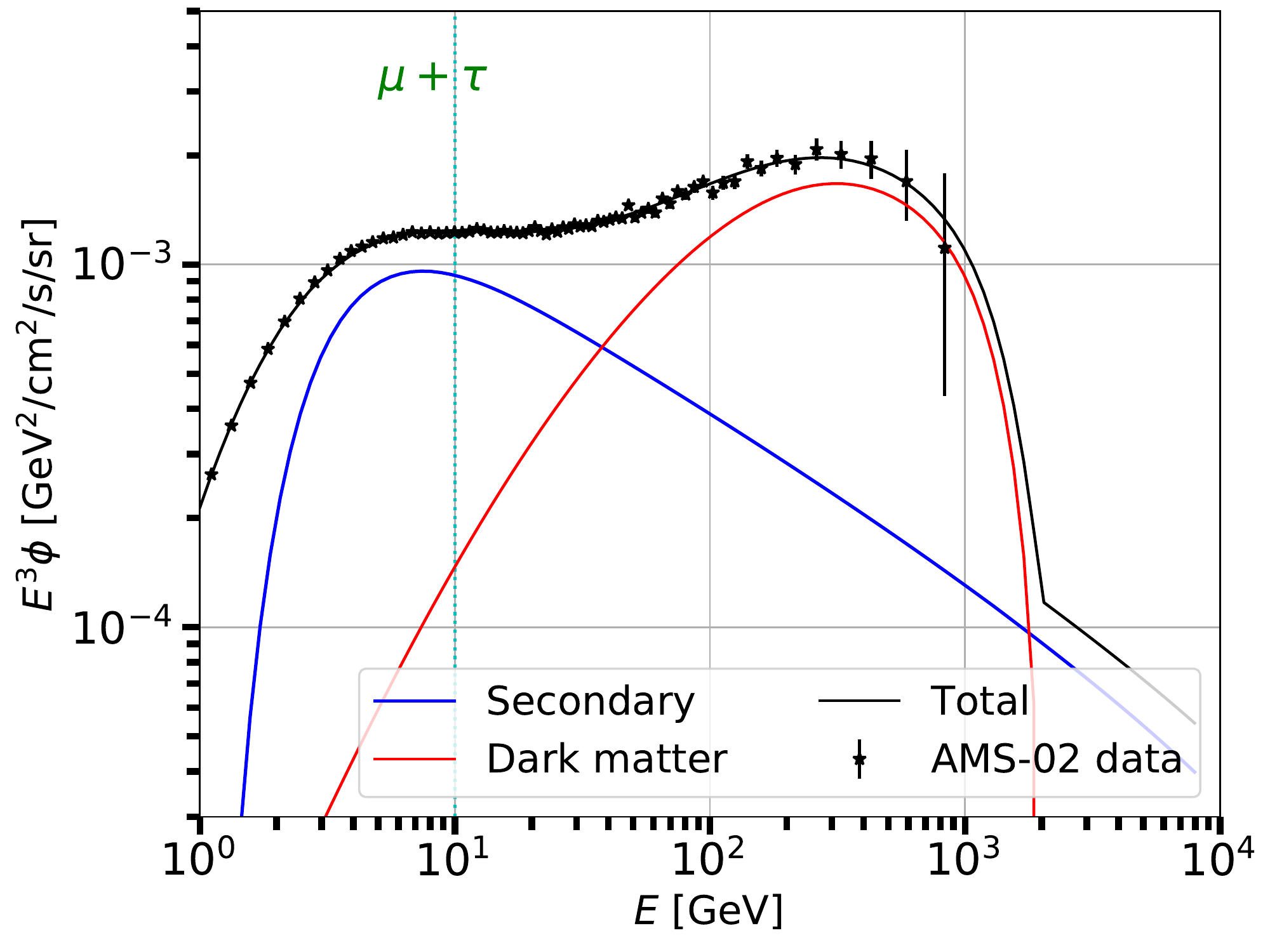}
    
  \end{subfigure}
    \caption{  Results of our fit on the AMS-02 data for the positron flux considering mixed channels.  }
    \label{123456}
\end{figure}

Fig.~\ref{123456} represents the results for the best-fit obtained considering mixed channels including $b\bar{b} \, \tau^+\tau^-$ , $\mu^+\mu^- \, b\bar{b}$ and $\mu^+\mu^- \, \tau^+\tau^-$. The best fit obtained is $\mu^+\mu^- \, b\bar{b}$ channel with a reduced $\chi^{2}$ of 0.944. For $\mu^+\mu^- \, b\bar{b}$ channel, the branching ratio for $\mu^+\mu^-$ is 0.427. The degree of freedom for our fit is 43, with an extra fit parameter. For the other 2 channels, the $\tau^+\tau^-$ channel dominates the branching ration in both mixed channels with $Br\sim 1$. For all the 3 mixed channels we obtain a better fit (in comparison with reduced $\chi^{2}$ value) than pure channels alone.
It is not possible to compare the data of mixed channels with other data as there is no report on the limit of DM cross-section for mixed channels.

\begin{table}[h!]
\centering
\begin{tabular}{ |p{3cm}||p{3cm}|p{3cm}|p{3cm}|  }
 \hline
 \multicolumn{4}{|c|}{Fit parameters for different mixed channels} \\
 \hline
 Par./channels $\chi^2$& $b + \tau$& $\mu +b$ & $\mu + \tau$\\
 \hline
 $M$ [GeV]& 1989.5 & 673.0 & 1995\\
 $\langle \sigma v \rangle$ [cm$^3$/s] & $3.39 \times 10^{-23}$& $8.91 \times 10^{-24}$& $3.16 \times 10^{-23}$\\
 $q$ & 1.16 & 0.700 & 1.18\\
 $Br$ & 0.98 ($\tau$) & 0.427 ($\mu$) & 0 ($\mu$)\\
 $\tilde{\chi}^2$ & 0.65 &1.54  & 0.65\\
\hline
 
\end{tabular}
\caption{Best-fit parameters and reduced $\chi^2$ for 3 different mixed channels $b + \tau, \mu +b$ and $\mu + \tau$.}
\label{table2}
\end{table}

\section{Conclusions} \label{conclution}

In this paper we perform a fit to the recently published AMS-02 positron data \cite{aguilar2021alpha,aguilar2019towards,aguilar2024towards} using the secondary production and a primary component that comes from DM annihilation from the Milky Way halo. 
We assume pure channels where DM annihilates into a couple of SM particles. We also test mixed channels when DM annihilates into two different channels with different branching ratios. The channel that fits the best data is obtained for DM annihilating into
$\tau^+ \tau^-$ with a reduced $\chi^{2}$ of 0.64, a DM mass of about 2 TeV and $\langle \sigma v \rangle = 3 \times 10^{-23}$ cm$^3$/s. 
The $b\bar{b}$ ($\mu^+\mu^-$) DM annihilation channel gives also a good fit with $M=43$ ($0.7$) TeV and $\langle \sigma v \rangle = 1 \times 10^{-21}$ ($4 \times 10^{-24}$) cm$^3$/s.
As for the mixed channels, the best fit was obtained with $\tau^+\tau^-$ with $b\bar{b}$ channel with a reduced $\chi^{2}$ of 0.65. For $\tau + b$ channel, The Branching ratio for $\tau$ is 0.98. 

In the second part of the paper, we compare the best-fit values of DM mass and $\langle \sigma v \rangle$ with the published upper limits from HESS for an analysis of the Galactic center \cite{abdallah2020search} and Fermi-LAT data for a combined search towards dSphs \cite{albert2020combined}. The only channel that fits well positron AMS-02 data and is compatible with $\gamma$-ray upper limits is the $\mu^+\mu^-$ while the other 2 are excluded by the limits obtained with HESS and Fermi-LAT data. 
In particular, we concluded that DM annihilating into $b\bar{b}$ and into $\tau^+\tau^-$ can contribute at most at the few $\%$ and a few tens of $\%$ to the positron data.
Other primary sources of positrons are not considered in our analysis as the conservative limits for DM are the focus of the analysis.  positron production from DM self- annihilation is a promising source for unexpected positron excess for AMS-02 data.

\bibliographystyle{plain}
\bibliography{main}
\end{document}